\documentstyle[12pt,caption]{article}
\newcommand {\ignore}[1]{}

\newcommand{\bc}{\begin{center}}
\newcommand{\ec}{\end{center}}

\def\ifmath#1{\relax\ifmmode #1\else $#1$\fi}
\def\3quarter{{\textstyle{3 \over 4}}}

\def\ra{\rightarrow}

\def\lf{\leaders\hbox to 1em{\hss.\hss}\hfill}
\def\ZP{$Z^\prime$ }
\def\21{$SU(2) \ot U(1)$}
\def\321{$SU(3) \ot SU(2) \ot U(1)$}
\def\ne{\hbox{$\nu_e$ }}
\def\nm{\hbox{$\nu_\mu$ }}
\def\nt{\hbox{$\nu_\tau$ }}
\def\ns{\hbox{$\nu_{sterile}$ }}

\def\Nt{\hbox{$N_\tau$ }}
\def\ns{\hbox{$\nu_S$ }}

\def\O{\hbox{$\cal O$ }}

\def\mnt{\hbox{$m_{\nu_\tau}$ }}

        \def\etc{\hbox{\it etc. }}
\def\eg{\hbox{\it e.g., }}        
\def\etal{\hbox{\it et al., }}
\def\rh{\hbox{right-handed }}

\def\gau{\hbox{gauge }}

\def\neu{\hbox{neutrino }}
\def\sa{\hbox{such as }}

\def\neus{\hbox{neutrinos }}

\def\eq#1{{eq. (\ref{#1})}}

\def\lsim{\raise0.3ex\hbox{$\;<$\kern-0.75em\raise-1.1ex\hbox{$\sim\;$}}}
\def\gsim{\raise0.3ex\hbox{$\;>$\kern-0.75em\raise-1.1ex\hbox{$\sim\;$}}}
\def\bel{\begin{letter}}
\def\eel{\end{letter}}
\def\beq{\begin{equation}}
\def\eeq{\end{equation}}
\def\bef{\begin{figure}}
\def\eef{\end{figure}}
\def\bet{\begin{table}}
\def\eet{\end{table}}
\def\bea{\begin{eqnarray}}
\def\ba{\begin{array}}
\def\ea{\end{array}}
\def\bi{\begin{itemize}}
\def\ei{\end{itemize}}
\def\ben{\begin{enumerate}}
\def\een{\end{enumerate}}
\def\ra{\rightarrow}
\def\ot{\otimes}
\def\eea{\end{eqnarray}}
\def\apj#1#2#3{          {\it Astrophys. J. }{\bf #1} (19#2) #3}

\def\jel#1#2#3{         {\it Journal Europhys. Lett. }{\bf #1} (19#2) #3}

\def\ib#1#2#3{           {\it ibid. }{\bf #1} (19#2) #3}
\def\nat#1#2#3{          {\it Nature }{\bf #1} (19#2) #3}
\def\nps#1#2#3{          {\it Nucl. Phys. B (Proc. Suppl.) }
                         {\bf #1} (19#2) #3}
\def\np#1#2#3{           {\it Nucl. Phys. }{\bf #1} (19#2) #3}
\def\pl#1#2#3{           {\it Phys. Lett. }{\bf #1} (19#2) #3}
\def\pr#1#2#3{           {\it Phys. Rev. }{\bf #1} (19#2) #3}
\def\prep#1#2#3{         {\it Phys. Rep. }{\bf #1} (19#2) #3}
\def\prl#1#2#3{          {\it Phys. Rev. Lett. }{\bf #1} (19#2) #3}
\def\pw#1#2#3{          {\it Particle World }{\bf #1} (19#2) #3}

\def\zp#1#2#3{           {\it Zeit. fur Physik }{\bf #1} (19#2) #3}

\def\n.c.#1#2#3{         {\it Nuovo Cim. }{\bf #1} (19#2) #3}
\def\r.n.c.#1#2#3{       {\it Riv. del Nuovo Cim. }{\bf #1} (19#2) #3}
\def\sjnp#1#2#3{         {\it Sov. J. Nucl. Phys. }{\bf #1} (19#2) #3}

\def\mpl#1#2#3{          {\it Mod. Phys. Lett. }{\bf #1} (19#2) #3}

\def\ppnp#1#2#3{           {\it Prog. Part. Nucl. Phys. }{\bf #1} (19#2) #3}

\def\tp{these proceedings}

\def\opc{\hbox{{\sl op. cit.} }}

\relax
\begin{document}
\begin{center}

\newcommand{\dis}{\displaystyle}

{\Large \bf SIGNATURES FROM PHYSICS BEYOND THE STANDARD MODEL}
\vskip 1cm
{\large {\bf Jos\'e W. F. Valle \\}
E-mail VALLE at vm.ci.uv.es or 16444::VALLE}
\footnote{Work supported by DGICYT under grant PB92-0084 }

\vskip 1cm

{\sl Instituto de F\'{\i}sica Corpuscular - C.S.I.C.,
Departament de F\'{\i}sica Te\`orica, Universitat de Val\`encia\\
46100 Burjassot, Val\`encia, SPAIN         }\\

\end{center}

\begin{abstract}

A brief review is made of some of the experimental signatures that may
be associated to a certain class of extensions of the standard model.
The material of these lectures is divided into two sections.

After briefly sketching the present observational status
of the neutrino masses I consider various schemes of neutrino
mass generation, including those which are motivated by
present experimental hints from solar and atmospheric
neutrinos, as well as cosmological data on the
amplitude of primordial density fluctuations.

Then some of the physics motivations and potential of
various extensions of the standard model related to the
electroweak breaking sector, such as supersymmetry, and
extensions of the gauge boson sector are reviewed.

The new signatures associated with both types of extension
may all be accessible to experiments performed either at
accelerators or at underground installations. The
complementarity between these two approaches in the
search for signals beyond the standard model is most
vividly manifest in the field of neutrino physics.

\end{abstract}

\section{Introduction}

Although extremely successful wherever it has
been tested, our present standard \21 model leaves
open many of the fundamental issues of present-day
particle physics. In the flavour sector, the most fundamental
problems involve the understanding of what lies behind
the mechanism of mass generation in general, as well
as the properties of neutrinos.

As is well known, the standard model relies on the Higgs
mechanism which implies the existence of a fundamental scalar
boson. If an elementary higgs boson exists it is widely believed
that some stabilizing principle - e.g. supersymmetry (SUSY) -
should be operative at the electroweak scale in order
to explain the stability of its mass scale against quantum
corrections associated with physics at superhigh energies.
The observed joining of the three \gau coupling strengths
as they are evolved from the presently accessible energies
up to to a common scale of $\sim 10^{16}$ GeV provides
circumstantial evidence that SUSY does indeed seem to set
in somewhere at $M_{SUSY} \sim 10^3$ GeV. Unveiling the
details of this rich structure constitutes one of the main
goals in the agenda of the next generation of elementary
particle colliders.

Another fundamental question mark in the standard model
refers to the properties of neutrinos. Apart from being a
theoretical puzzle, in the sense that there is no principle
that dictates that neutrinos are massless, as postulated in
the standard model, nonzero masses may in fact be required
in order to account for a natural explanation of the data
on solar and atmospheric neutrinos, as well as for the hot
dark matter component of the universe. The implications
of detecting nonzero neutrino masses could be very
far reaching for the understanding of fundamental
issues in particle physics, astrophysics, as well
as the structure of our universe.

These two different types of extensions may be related
in some models. As an example, I will consider the case of
supersymmetric models with spontaneously broken R parity,
which necessarily imply nonvanishing neutrino masses.
I will describe how the extensions of the basic picture
that seek to address the above two issues, such as higher
unification and supersymmetry, may lead to extensions
of the lepton and/or Higgs boson multiplet content, and
thereby affect the physics of the electroweak sector in
an important way. How to probe this physics both in
accelerator as well as underground experiments
will also be described.

\section{Neutrino Mass}

Neutrinos are the only apparently massless electrically neutral
fermions in the standard model and the only ones without \rh partners.
It is rather mysterious that they seem to be so special
when compared with the other fundamental fermions.
Indeed, having no electric charge, a majorana mass
term for neutrinos may arise even in the absence of \rh
components. However, many unified extensions of the
standard model, such as SO(10), do require the presence
of \rh neutrinos in order to realize the extra symmetry.
Either way one expects \neus to be massive. Moreover,
there is, in these theories, a natural mechanism,
called {\sl seesaw}, to understand the relative smallness
of \neu masses \cite{GRS,fae}. In general the seesaw
mechanism provides just a general scheme, rather than
detailed predictions. These will depend, among other
factors, upon the structure not only of the Dirac type
entries, but also on the possible texture of the large
Majorana mass term \cite{Smirnov}.

Although attractive, the seesaw mechanism is by no means
the only way to generate \neu masses. There are many other
attractive possibilities, some of which do not require any
new large mass scale. The extra particles required to
generate the \neu masses have masses at scales accessible
to present experiments \cite{zee.Babu88}.

It is also quite plausible that B-L or lepton number,
instead of being part of the gauge symmetry \cite{LR}
may be a spontaneously broken global symmetry. The scale
at which such a symmetry gets broken does not need to high,
as in the original proposal \cite{CMP}, but can be rather
low, close to the weak scale \cite{JoshipuraValle92}.
Such a low scale for lepton number breaking could have
important implications not only in astrophysics and
cosmology but also in particle physics.

This large diversity of possible schemes and the
lack of a theory for the Yukawa couplings imply
that present theory is not capable of predicting the
scale of \neu masses any better than it can fix the masses
of the other fermions, like that of the muon. As a
result one should at this point turn to experiment.

\subsection{Limits}

There are several limits on \neu masses that follow
from observation. The laboratory bounds may be
summarized as \cite{PDG94}
\beq
\label{1}
m_{\nu_e} 	\lsim 5 \: \rm{eV}, \:\:\:\:\:
m_{\nu_\mu}	\lsim 250 \: \rm{keV}, \:\:\:\:\:
m_{\nu_\tau}	\lsim 31  \: \rm{MeV}
\eeq
and follow purely from kinematics. These are the
most model-independent of the \neu mass limits.
The improved limit on the \ne mass from beta decays
was recently given by Lobashev \cite{Erice}, while that
on the \nt mass may be substantially improved at a future tau
factory \cite{jj}.

In addition, there are limits on neutrino masses
that follow from the nonobservation of neutrino
oscillations \cite{granadaosc}. They involve
\neu mass differences versus mixing, and disappear
in the limit of unmixed neutrinos. The present
situation as well as future prospects to probe
for neutrino oscillation parameters at long baseline
experiments is given in Figure \ref{long}.
\bef
\vspace{8.4cm}
\caption{
Oscillation parameters probed at present and future
neutrino experiments}
\label{long}
\eef

Another important limit arises from the
non-observation of ${\beta \beta}_{0\nu}$ decay, i.e.
the process by which nucleus $(A,Z-2)$ decays to
$(A,Z) + 2 \ e^-$. This lepton number violating
process would arise from majorana \neu exchange.
In fact, as shown in ref. \cite{BOX}, a nonvanishing
${\beta \beta}_{0\nu}$ decay rate requires \neus to be
majorana particles, irrespective of which mechanism
induces it. This establishes a very deep connection
which, in some special models, may be translated
into a lower limit on the \neu masses.
The negative searches for ${\beta \beta}_{0\nu}$
in $^{76} \rm{Ge}$ and other nuclei leads to a limit of
about one or two eV \cite{Avignone}
on a weighted average \neu mass parameter
characterizing this process. Better sensitivity is expected
from the upcoming enriched germanium experiments.
Although rather stringent, this limit
may allow relatively large \neu masses, as there
may be strong cancellations between the contributions of
different neutrino types. This happens automatically
in the case of a Dirac \neu due to the lepton number
symmetry \cite{QDN}.

In addition to laboratory limits, there is a cosmological
bound that follows from avoiding the overabundance of
relic neutrinos \cite{KT}
\beq
\sum_i m_{\nu_i} \lsim 50 \: \rm{eV}
\label{rho1}
\eeq
This limit only holds if \neus are stable on cosmological
time scales. There are many models where neutrinos decay
into a lighter \neu plus a majoron \cite{fae},
\beq
\nu_\tau \ra \nu_\mu + J \:\: .
\label{NUJ}
\eeq
Lifetime estimates in various majoron models have
been discussed in ref. \cite{V}. These decays can
be fast enough to obey the cosmological limits coming
from the critical density requirement, as well as those
that come from primordial big-bang nucleosynthesis
\cite{BBNUTAU}. Note also that, since these decays
are $invisible$, they are consistent with all
astrophysical observations.
In view of the above it is worthwhile to continue
in the efforts to improve present laboratory \neu mass
limits, including searches for distortions in the energy
distribution of the electrons and muons coming from weak
decays \sa $\pi, K \ra e \nu$, $\pi, K \ra \mu \nu$, as
well as kinks in nuclear $\beta$ decays \cite{Deutsch}.

In addition to the above limits there are
some positive {\sl hints} for neutrino masses
that follow from the following cosmological,
astrophysical and laboratory observations.

\subsection{Dark Matter}

Recent observations of cosmic background temperature
anisotropies on large scales by the COBE  satellite
\cite{cobe}, when combined with cluster-cluster correlation
data e.g. from IRAS \cite{iras}, indicate the need for the existence
of a hot {\sl dark matter} component, contributing
about 30\% to the total mass density \cite{cobe2}.
A good fit is provided by a massive neutrino, for
example, a tau neutrino in the few eV mass range.
This suggests the possibility
of having observable \ne to \nt or \nm to \nt
oscillations that may be accessible to the CHORUS
and NOMAD experiments at CERN, as well as at the
proposed P803 experiment at Fermilab \cite{chorus}.
This mass scale is also consistent with the recent
preliminary hints in favour of neutrino oscillations
recently reported by the LSND experiment \cite{Caldwell}.

\subsection{Solar Neutrinos}

The data collected up to now by Homestake and Kamiokande,
as well as by the low-energy data on pp neutrinos from
the GALLEX and SAGE experiments still pose a persisting
puzzle \cite{Davis,granadasol}.
Comparing the data of GALLEX with the Kamiokande data
indicates the need for a reduction of the $^7 $ Be flux
relative to the standard solar model expectation. Inclusion
of the Homestake data only aggravates the discrepancy,
suggesting that the solar \neu problem is indeed a real problem.
The allowed one sigma region for $^7 $ Be and $^8$ Be
fluxes is obtained as the intersection of the region to
the left of line labelled 91 with the region labelled KAMIOKA
in Figure \ref{solardata}.
The lines are normalized with respect to the reference
solar model of Bahcall and collaborators. Including
the Homestake data of course only aggravates the
discrepancy as can be seen from Figure \ref{solardata}.
\bef
\vspace{9.1cm}
\caption{
Allowed one sigma bands for $^7 $ Be and $^8$ Be fluxes
from all solar neutrino data}
\label{solardata}
\eef

Thus if one takes all data simultaneously one concludes
that the simplest astrophysical solutions to the solar \neu data
are highly disfavored and that one needs new physics in the \neu
sector to account for the data \cite{NEEDNEWPHYSICS}.
The most attractive possibility is to assume the
existence of \neu conversions involving very small
\neu masses $\sim 10^{-3}$ eV \cite{MSW}.
The region of parameters allowed by present
experiments is given in ref. \cite{Hata.MSWPLOT}.
Note that the fits favour the non-adiabatic over the
large mixing solution, due mostly to the larger reduction
of the $^7 $ Be flux found in the former.

\bef
\vspace{9.1cm}
\caption{Region of solar \neu oscillation parameters
allowed by experiment}
\label{msw}
\eef

\subsection{Atmospheric Neutrinos}

An apparent decrease in the expected flux of atmospheric
$\nu_\mu$'s relative to $\nu_e$'s arising from the decays
of $\pi$'s, $K$'s and secondary muon decays produced in
the atmosphere, has been observed in two underground
experiments, Kamiokande and IMB, and possibly also at
Soudan2 \cite{atm}. Although the predicted absolute
fluxes of \neus produced by cosmic-ray interactions in the
atmosphere are uncertain at the 20 \% level, their
ratios are expected to be accurate to within 5 \%.

This atmospheric neutrino deficit can be ascribed to
\neu oscillations.
Combining these experimental results with observations
of upward going muons made by Kamiokande, IMB and Baksan,
and with the negative Frejus and NUSEX results \cite{up}
leads to the following range of neutrino oscillation
parameters
\beq
\label{atm0}
\Delta m^2_{\mu \tau} \approx 0.005 \: - \: 0.5\ \rm{eV}^2,\
\sin^22\theta_{\mu \tau} \approx 0.5
\eeq
Recent results from Kamiokande on higher energy \neus
strengthen the case for an atmospheric \neu problem
\cite{atm1} as shown in Figure \ref{kamglasgow}.
\bef
\vspace{9.1cm}
\caption{Region of atmospheric \neu oscillation parameters
from recent Kamiokande data.}
\label{kamglasgow}
\eef

\subsection{Models Reconciling Present Hints.}

Can we reconcile the present hints from astrophysics and
cosmology in the framework of a consistent elementary
particle physics theory? The above observations suggest
an interesting theoretical puzzle whose possible
resolutions will now be discussed.

\subsubsection{Three Almost Degenerate Neutrinos}

It is difficult to reconcile these three observations
simultaneously in the framework of the simplest seesaw model
with just the three known \neus. The only possibility to fit
these observation in a world with just the three neutrinos
of the standard model is if all of them have nearly the
same mass $\sim$ 2 eV \cite{caldwell}.

It is known that the general seesaw models have
two independent terms giving rise to the light neutrino masses.
The first is an effective triplet vacuum expectation value
\cite{2227} which is expected to be small in left-right
symmetric models \cite{LR}. Based on this fact one can
in fact construct extended seesaw models where the main
contribution to the light \neu masses ($\sim$ 2 eV) is universal,
due to a suitable horizontal symmetry, while the splittings
between \ne and \nm explain the solar \neu deficit and that
between \nm and \nt explain the atmospheric \neu anomaly \cite{DEG}.

\subsubsection{Three Active plus One Sterile Neutrino}

The alternative way to fit all the data is to add a
fourth \neu species which, from the LEP data on the
invisible Z width, we know must be of the sterile type,
call it \ns. The first scheme of this type gives mass
to only one of the three neutrinos at the tree level,
keeping the other two massless \cite{OLD}.
In a seesaw scheme with broken lepton number, radiative
corrections involving gauge boson exchanges will give
small masses to the other two neutrinos \ne and \nm
\cite{Choudhury}. However, since the singlet \neu is
superheavy in this case, there is no room to account
for the three hints discussed above.

Two basic schemes have been suggested to keep the sterile
neutrino light due to a special symmetry. In addition to the
sterile \neu \ns, they invoke additional Higgs bosons beyond
that of the standard model, in order to generate radiatively
the scales required for the solar and atmospheric \neu
conversions. In these models the \ns either lies at the dark matter
scale \cite{DARK92} or, alternatively, at the solar \neu scale
\cite{DARK92B}.
In the first case the atmospheric
\neu puzzle is explained by \nm to \ns oscillations,
while in the second it is explained by \nm to \nt
oscillations. Correspondingly, the deficit of
solar \neus is explained in the first case
by \ne to \nt oscillations, while in the second
it is explained by \ne to \ns oscillations. In both
cases it is possible to fit all observations together.
However, in the first case there is a clash with the
bounds from big-bang nucleosynthesis. In the latter
case the \ns is at the MSW scale so that nucleosynthesis
limits are satisfied. They single out the nonadiabatic
solution uniquely. Note however that, since the
mixing angle characterizing the \nm to \nt
oscillations is nearly maximal, the second
solution is in apparent conflict with \eq{atm0}
but agrees with Figure \ref{kamglasgow}, taken from
ref. \cite{atm1}. Moreover,
it can naturally fit the recent preliminary hints of
neutrino oscillations of the LSND experiment \cite{Caldwell}.

Another theoretical possibility is that all active
\neus are very light, while the sterile \neu \ns is
the single \neu responsible for the dark matter
\cite{DARK92D}.

\subsection{New Signatures in the Lepton Sector.}

There are many motivations to extend the lepton sector
of the electroweak theory. Extra heavy leptons may arise
in models with a higher unification, for example those
with left-right symmetry, SO(10) grand unified models,
or superstrings. These models may contain isosinglet
neutral heavy leptons and typically, also neutrino
masses \cite{fae}.

These isosinglet neutral heavy leptons (NHLS) may induce
lepton flavour violating (LFV) decays \sa
$\mu \rightarrow e \gamma$, which are exactly forbidden
in the standard model. Although these are a generic
feature of models with massive \neus, in some cases,
they may proceed in models where \neus are strictly
massless \cite{SST,CP,BER}.

In the simplest models of seesaw type \cite{GRS} the NHLS are
superheavy so that the expected rate for LFV processes
is expected to be low, due to limits on \neu masses.
However, in other variants \cite{SST} this is not the case
\cite{BER,CP} and this suppression need not be present.
Indeed, present constraints on weak universality violation
allow for decay branching ratios larger than the present
experimental limits \cite{3E} so that these already
are probing the masses and admixtures of the
NHLS with considerable sensitivity. Similar estimates
can be done for the corresponding tau decays \cite{3E,Pila}.
The results are summarized in Table 1. As an illustration,
Figure \ref{pila} gives the expectations for the three
charged lepton decays of the tau, taken from ref. \cite{Pila}.
\bef
\vspace{9.1cm}
\caption{Expected branching ratios for $\tau \ra 3e$
(solid) and $\tau \ra \mu \mu e$}
\label{pila}
\eef
Clearly these branching ratios lie within the
sensitivities of the planned tau and B factories,
as shown in ref. \cite{TTTAU}.
\begin{table}
\begin{center}
\caption{Allowed $\tau$ decay branching ratios }.
\begin{displaymath}
\begin{array}{|c|cr|}
\hline
\mbox{channel} & \mbox{strength} & \mbox{} \\
\hline
\tau \rightarrow e \gamma ,\mu \gamma &  \lsim 10^{-6} & \\
\tau \rightarrow e \pi^0 ,\mu \pi^0 &  \lsim 10^{-6} & \\
\tau \rightarrow e \eta^0 ,\mu \eta^0 &  \lsim 10^{-6} - 10^{-7} & \\
\tau \rightarrow 3e , 3 \mu , \mu \mu e, \etc &  \lsim 10^{-6} - 10^{-7} & \\
\hline
\end{array}
\end{displaymath}
\end{center}
\end{table}

The physics of rare $Z$ decays nicely complements what
can be learned from the study of rare LFV muon and tau decays.
The stringent limits on $\mu \rightarrow e \gamma$ preclude any
possible detectability at LEP of the corresponding
$Z \rightarrow e \mu$ decay. However the decays with
tau number violation, $Z \ra e\tau$ or $\mu\tau$ can be large.
Similar statements can be made also for the CP violating Z decay
asymmetries in these LFV processes \cite{CP}. Under realistic
luminosity and experimental resolution assumptions, however, it is
unlikely that one will be able to see these decays of the Z at
LEP without a high luminosity option \cite{ETAU}.
In any case, there have been dedicated experimental
searches which have set good limits \cite{opal}.

If the NHLS are lighter than the $Z$, they may also be
produced directly in Z decays such as
\footnote{There may also be CP violation in lepton
sector, even when the known \neus are strictly massless
and lead to Z decay asymmetries \O($10^{-7}$) \cite{CP}} \cite{CERN},
\begin{equation}
Z \rightarrow N_{\tau} + \nu_{\tau}
\end{equation}
Note that the isosinglet neutral heavy lepton
\Nt is singly produced, through the off-diagonal
neutral currents characteristic of models containing
doublet and singlet leptons \cite{2227}.
Subsequent \Nt decays would then give rise to
large missing energy events, called zen-events.
Expectations for the attainable rates for such
processes are illustrated in Figure \ref{cern}, taken
from ref. \cite{CERN}
\bef
\vspace{9.1cm}
\caption{LEP sensitivities to $Z  \ra N \nu$ decays}
\label{cern}
\eef
One sees that this branching ratio can be
as large as $\lsim 10^{-3}$ a value that is already
superseded by the good limits on such decays from
the searches for acoplanar jets and lepton pairs from $Z$
decays at LEP, although some inconclusive hints have
been recently reported by ALEPH \cite{opal}
\begin{table}
\begin{center}
\caption{Allowed branching ratios for rare $Z$
decays. }
\begin{displaymath}
\begin{array}{|c|cr|}
\hline
\mbox{channel} & \mbox{strength} & \mbox{} \\
\hline
Z \rightarrow \Nt \nt &  \lsim 10^{-3} & \\
Z \rightarrow e \tau &  \lsim 10^{-6} - 10^{-7} & \\
Z \rightarrow \mu \tau &  \lsim 10^{-7} & \\
\hline
\end{array}
\end{displaymath}
\end{center}
\end{table}

Finally we note that there can also be large rates for
lepton flavour violating decays in models with radiative
mass generation \cite{zee.Babu88}. For example, this is
the case in the models proposed to reconcile present
hints for \neu masses \cite{DARK92}. The expected decay
rates may easily lie within the present experimental
sensitivities and the situation should improve at PSI
or at the proposed tau-charm factories.

\subsection{Outlook}

Besides being suggested by theory, neutrino masses
seem to be required to fit present astrophysical and
cosmological observations, in addition to the recent
LSND hints \cite{Caldwell}.

Neutrinos could be responsible for a wide variety of
measurable implications at the laboratory. These new
phenomena would cover an impressive range of energies,
starting with $\beta$ and nuclear $\beta \beta_{0\nu}$
decays. Searches for the latter with enriched germanium
could test the quasidegenerate neutrino scenario for
the joint explanation of hot dark matter and
solar and atmospheric \neu anomalies.  Moving to
neutrino oscillations, here one expects much larger
regions of oscillation parameters in the \ne to \nt
and \nm to \nt channels will be be probed by the
accelerator experiments at CERN than now possible
with present accelerators and reactors.
On the other hand more data from low energy pp
neutrinos as well as from Superkamiokande, Borexino,
and Sudbury will shed light on the solar neutrino issue.
Fortunately these experiments are expected to run in the
next couple of years or so.

For the far future we look forward to the possibility
of probing those regions of \nm to \ne or \ns oscillation
parameters suggested by present atmospheric \neu data.
This will be possible at the next generation of long
baseline experiments.
Similarly, a new generation of experiments capable
of more accurately measuring the cosmological
temperature anisotropies at smaller angular scales than
COBE, would test different models of structure formation,
and presumably shed further light on the need for hot
\neu dark matter.

Neutrinos may also imply rare processes with lepton
flavour violation, as well as new signatures at LEP
energies and even higher, whose allowed rates have
been summarized in Tables 1 and 2. Such experiments
are complementary to those at low energies and can
also indirectly test \neu properties in an important way.

\section{ Electroweak Symmetry Breaking}

A lot of research effort has been recently devoted to
the physics associated to the electroweak breaking sector
and its possible manifestations at present and future
particle colliders. If indeed the higgs boson exists
as an elementary particle, the forerunner in these
investigations is the study of supersymmetric extensions
of the standard model and its corresponding experimental
searches at high energy accelerators.

The prototype of these models is called the minimal
supersymmetric standard model (MSSM) \cite{mssm}. This
model realizes SUSY in the presence of a discrete R parity
($R_p$) symmetry, postulated {\sl ad hoc}. Under this symmetry
all standard model particles are even while their partners
are odd. As a result of this selection rule, in
the so-called minimal supersymmetric standard model
SUSY particles are only produced in pairs, with the
lightest of them (LSP) being stable. It has been
suggested as a candidate for the cold dark matter of the
universe and several methods of detection at underground
installations have been suggested \cite{Bernabei}.

So far all searches for supersymmetric particles have
been negative. However, presently accessible energies
cover only a small part of the parameter space of
supersymmetric theories. One may summarize the
present situation as follows. The electrically
charged weakly interacting SUSY states, sleptons,
charginos as well as SUSY higgs bosons have bounds
close to the available beam energy at LEP. There
is only a small room for improvement left at LEP1
on the masses of the electrically neutral SUSY
particles. As for the strongly interacting SUSY states,
gluinos and squarks, their mass bounds come form
the Tevatron and there is little room for improvement
with the present setup.

Thus it seems that one has to wait for the new generation
of particle colliders, LEP2 and the LHC in order to
improve the search potential for supersymmetric models.
Indeed, this topic forms one of the important goals
in the agenda of these elementary particle colliders.

As will be shown in the next section, one has not yet
reached the border of what can be reached with present
installations in the searches for SUSY particles if one
abandons the assumption that R parity is conserved.
Indeed one can have genuine SUSY signals that can be
searched for even at LEP1 with the required sensitivity
to make the searches meaningful.

\subsection{Supersymmetry.}

Unfortunately there is no clue as to how SUSY is
realized. Nobody knows the origin of the R parity
symmetry and whether it is indeed a necessary requirement
to impose on supersymmetric extensions of the standard model.
Therefore there is no firm theoretical basis for
the most popular $ansatz$, the minimal supersymmetric
standard model (MSSM). It is indeed of great interest
to investigate theories without R parity \cite{fae}.

There are many ways to break it, either
explicitly or spontaneously (RPSUSY models).
Here we focus on the case of spontaneous $R_p$
breaking in the \21 theory. The viability of this
possibility has been recently demonstrated.
The breaking of R-parity is driven by right-handed
{\sl isosinglet} sneutrino vacuum expectation values
(VEVS) \cite{MASI_pot3}, so that the associated Goldstone
boson (majoron) is mostly singlet and as a result
the $Z$ does not decay by majoron emission, in
agreement with LEP observations \cite{LEP1}.

If R parity is broken spontaneously it shows up
primarily in the couplings of the W and the Z,
leading to rare $Z$ decays such as the single
production of the charginos and neutralinos
\cite{ROMA}, for example,
\begin{equation}
Z \rightarrow \chi^{\pm} \tau^{\mp}
\end{equation}
where the lightest chargino mass is assumed to be smaller than
the Z mass. In the simplest models, the magnitude of R parity
violation is correlated with the nonzero value of the \nt mass
and is restricted by a variety of experiments. Nevertheless
the R parity violating Z decay branching ratios, as an example,
can easily exceed $10^{-5}$, well within present LEP sensitivities.
This is illustrated in Figure \ref{chitau}.
\bef
\vspace{9.1cm}
\caption{Allowed branching ratios for $Z \ra \chi^{\pm} \tau^{\mp}$}
\label{chitau}
\eef
Similarly, the lightest neutralino (LSP) could also be
singly-produced as \cite{ROMA}
\beq
Z \rightarrow \chi^0 \nu_\tau
\eeq
Being unstable due to R parity violation, $\chi^0$ is
not necessarily an origin of events with missing energy,
since some of its decays are into charged particles.
Thus the decay $Z \rightarrow \chi^0 \nu_\tau$ would give rise to
zen events, similar to those of the MSSM but where the missing
energy is carried by the \nt. Another possibility for zen events
in RPSUSY is the usual pair neutralino production process, where
one $\chi^0$ decays visibly and the other invisibly. The
corresponding zen-event rates can be larger than in the MSSM.

Although the \nt can be quite massive in these models,
it is perfectly consistent with cosmology \cite{KT}
including primordial nucleosynthesis \cite{BBNUTAU},
since it decays sufficiently fast by majoron emission
\cite{V}. On the other hand, the
\ne and \nm have a tiny mass difference in the model of
ref. \cite{MASI_pot3}. This mass difference can be chosen to lie
in the range where resonant \ne to \nm conversions provides
an explanation of solar \neu deficit \cite{MSW}. Due to this peculiar
hierarchical pattern, one can go even further, and regard the
rare R parity violating processes as a tool to probe the
physics underlying the solar \neu conversions in this model
\cite{RPMSW}. Indeed, the rates for such rare decays can be
used in order to discriminate between large and small mixing
angle MSW solutions to the solar \neu problem \cite{MSW}.
Typically, in the nonadiabatic region of small mixing one can have
larger rare decay branching ratios, as seen in Figure 5
of ref. \cite{RPMSW}.

It is also possible to find manifestations of
R parity violation at the superhigh energies available
at hadron supercolliders such as LHC. Either SUSY
particles, such as gluinos, are pair produced and
in their cascade decays the LSP decays or, alternatively,
one violates R parity by singly producing the SUSY states.
An example of this situation has been discussed in
ref. \cite{RPLHC}. In this reference one has
studied the single production of weakly interacting
supersymmetric fermions (charginos and neutralinos)
via the Drell Yan mechanism, leading to possibly
detectable signatures. More work on this will
be desirable.

Another possible signal of the RPSUSY models based
on the simplest \21 gauge group is rare decays of
muons and taus. In this model the spontaneous
violation of R parity generates a physical Goldstone
boson, called majoron. Its existence is quite consistent
with the measurements of the invisible $Z$ decay width at LEP,
as it is a singlet under the \21 \gau symmetry.
In this model the lepton number is broken close
to the weak scale and can produce a new class of
lepton flavour violating decays, such as those
with single majoron emission in $\mu$ and $\tau$
decays. These would be "seen" as bumps in the final
lepton energy spectrum, at half of the parent lepton
mass in its rest frame.
\begin{table}
\begin{center}
\caption{Allowed branching ratios for rare decays
in the RPSUSY model. $\chi$ denotes the lightest
electrically charged SUSY fermion (chargino)
and $\chi^0$ is the lightest neutralino.}
\begin{displaymath}
\begin{array}{|c|cr|}
\hline
\mbox{channel} & \mbox{strength} & \mbox{} \\
\hline
Z \rightarrow \chi \tau &  \lsim 6 \times 10^{-5} & \\
Z \rightarrow \chi^0 \nt &  \lsim 10^{-4} & \\
\hline
\tau \rightarrow \mu + J &  \lsim 10^{-3} & \\
\tau \rightarrow e + J &  \lsim 10^{-4} & \\
\hline
\end{array}
\end{displaymath}
\end{center}
\end{table}
The allowed rates for single majoron emitting $\mu$
and $\tau$ decays have been determined in ref. \cite{NPBTAU}
and are also shown in table 3 to be compatible with present
experimental sensitivities \cite{PDG94}. As an illustration,
I borrow Figure \ref{npbtau} from ref. \cite{NPBTAU}.
\bef
\vspace{9.1cm}
\caption{Allowed branching ratios for $\tau \ra e + J$ versus \mnt}
\label{npbtau}
\eef
This example also illustrates how the search for
rare decays can be a more sensitive probe of \neu
properties than the more direct searches for \neu
masses, and therefore complementary. Moreover, they are
ideally studied at a tau-charm factory \cite{TTTAU}.

\subsection{Higgs Bosons.}

Another possible, albeit quite indirect, manifestation
of the properties of \neus and the lepton sector is in
the electroweak breaking sector. Many extensions of the
lepton sector seek to give masses to \neus through the
spontaneous violation of an ungauged U(1) lepton number
symmetry, thus implying the existence of a physical
Goldstone boson, called majoron \cite{CMP}. As already
mentioned above this is consistent with the measurements
of the invisible $Z$ decay width at LEP if the majoron
is (mostly) a singlet under the \21 \gau symmetry.

Although the original majoron proposal was made
in the framework of the minimal seesaw model, and
required the introduction of a relatively high
energy scale associated to the mass of the \rh
\neus \cite{CMP}, there are many attractive
theoretical alternatives where lepton number
is violated spontaneously at the weak scale or
lower. In this case although the majoron has very
tiny couplings to matter and the \gau bosons, it
can have significant couplings to the Higgs bosons.
As a result one has the possibility that the Higgs
boson may decay with a substantial branching ratio
into the invisible mode \cite{JoshipuraValle92}
\begin{equation}
h \rightarrow J\;+\;J
\label{JJ}
\end{equation}
where $J$ denotes the majoron. The presence of
this invisible decay channel can affect the
corresponding Higgs mass bounds in an important way.

The production and subsequent decay of a Higgs boson
which may decay visibly or invisibly involves three independent
parameters: its mass $M_H$, its coupling strength to the Z,
normalized by that of the standard model, $\epsilon^2$, and its
invisible decay branching ratio.
The LEP searches for various exotic channels can be used
in order to determine the regions in parameter space
that are already ruled out, as described in ref.
\cite{alfonso}. The exclusion contour in the plane
$\epsilon^2$ vs. $M_H$, was shown in Figure \ref{alfonso2}
taken from ref. \cite{moriond}.

\bef
\vspace{9.1cm}
\caption{Region in the $\epsilon^2$ vs. $m_H$ that can be
excluded by the present LEP1 analyses (solid curve).
Also shown are the LEP2 extrapolations (dashed).}
\label{alfonso2}
\eef

Another mode of production of invisibly decaying
Higgs bosons is that in which a CP even Higgs boson
is produced at LEP in association with a massive
CP odd scalar \cite{HA}. This production mode is
present in all but the simplest majoron model
containing just one complex scalar singlet in addition
to the standard model Higgs doublet. Present limits on the
relevant parameters are given in Figure \ref{ha}, taken from
ref. \cite{HA}.  In this plot we have assumed
BR ($H \rightarrow J\:J$) = 100\% and a visibly
decaying A boson.
\bef
\vspace{9.1cm}
\caption{Limits on $\epsilon^2_{A}$ in the $m_A,m_H$ plane
that can be placed by present LEP1 searches based on the
 $e^+ e^- \rightarrow H \:A \rightarrow J\:J b\bar{b}$
production channel.      }
\label{ha}
\eef

Finally, the invisible decay of the Higgs boson may
also affect the strategies for searches at higher energies.
For example, the ranges of parameters that can be covered
by LEP2 searches for a total integrated luminosity of
500 pb$^{-1}$ and various centre-of-mass energies have
been given in Figure \ref{alfonso2}. Similar analysis were
made for the case of a high energy linear $e^+ e^-$ collider
(NLC) \cite{EE500}, as well as for the LHC \cite{granada}.

\subsection{New Gauge Bosons.}

Superstring extensions of the standard model suggest the
existence of additional gauge bosons at the TeV scale and
this may affect the lepton sector and the interactions of
neutrinos. Although there are other possibilities,
we focus here on models based on an underlying $E_6$
symmetry \cite{fae}.

The fantastic agreement found between the standard
model predictions and the experimental measurements
from the scale of the atom to that probed at LEP
places stringent restrictions on the existence of
an additional \ZP at low energies \cite{altarelli}.
Indeed, if such boson were sufficiently light and mixed with the
usual Z it would modify the couplings of leptons to the
Z and be thereby restricted by low energy neutral current
data, as well as by the LEP precision data on
Z decays \cite{altarelli}. In string models the Higgs sector
is constrained in such a way that these limits are strongly
correlated with the top quark mass \cite{B259}. This is
illustrated in Figure \ref{corr}, taken from ref. \cite{B259}.
\bef
\vspace{9.1cm}
\caption{Limits on \ZP bosons in constrained string type models
based on $E_6$}
\label{corr}
\eef
One sees that the recent data from the CDF collaboration leads
to constraints around a TeV on the \ZP mass for such string
type models based on the $E_6$ \gau group. The limits are
much weaker in the case of unconstrained models.

\subsection{Outlook}

There is a wealth of related phenomena covering a
broad range of energies and of experimental situations
that may probe the physics underlying the extensions
we have discussed here. They involve signatures
in the \neu sector, such as oscillations, neutrinoless
double beta decays and possible distortions in beta
decay spectra. A large number of related processes can
also manifest themselves at muon and tau factories and
at high energy $e^+e^-$ collisions (e.g. LEP and NLC).
These have been summarized in Table 3.
There are also good prospects to observe some of these
signatures at the upcoming hadron supercolliders LHC.
Examples of these processes range from $\mu$ and $\tau$
number violating decays, up the high energy processes
associated with the single production of SUSY fermions or
neutral heavy leptons (NHLS) at LEP or at a future hadron
supercollider. Finally let me highlight in this context
the rather peculiar possibility that the Higgs boson may
decay dominantly by two majoron emission, leading to missing
momentum events. As we saw, new search strategies are
required to cover this possibility. All of the above
effects related to nonstandard \neu properties may be
accessible to experiment.

\section*{Acknowledgements}
This paper has been supported by DGICYT under
Grant number PB92-0084. I thank the organizers for a very
pleasant meeting at Jaca. Special thanks are due to
Mercedes Fatas, for her charm and efficiency.

\bibliographystyle{ansrt}

\end{document}